 \documentstyle[12pt,twoside,fleqn,espcrc1,epsfig]{article} 

\newcommand{\psip}{\psi'}
\newcommand{\jpsi}{J/\psi}
\def\Journal#1&#2&#3(#4){#1{\bf #2}, #3 (#4)}

\def\NIMA{Nuclear Instruments and Methods {\bf A}}

\def\PLB{Physics Letters {\bf B}}
\def\PRL{Physical Review Letters }
\def\PRD{Physical Review {\bf D}}

\def\etal{et al.}

\newcommand{\AmS}{{\protect\the\textfont2
  A\kern-.1667em\lower.5ex\hbox{M}\kern-.125emS}}
\title{ Recent $\psi'$ Results at BES }

\author{F. Liu
\address{ On behalf of the BES Collaboration\\ Inst. of High Energy Physics, 
        P.O. Box 918(1), Beijing 100039}
        \thanks{E-mail: lfeng@hpws3.ihep.ac.cn, Web Site:  http://hpws1.ihep.ac.cn/$^\sim$lfeng}}
\begin{document}
\maketitle
\begin{abstract}
Based on $(3.79\pm0.31)\times10^6 ~\psip$ data sets collected with the BES 
detector at BEPC, the recent $\psip$, $\chi_{cJ}$ and $\eta_c$ results from BES 
are presented. Some results are compared with NRQCD.
\end{abstract}
\vskip 0.2cm 
\section {Studies of $\psip$ Decays }

Charmonium physics is always one of the interesting and intriguing
fields of particle physics. Charmonium provides us an excellent 
and simple system to study QCD, the  production and decay 
mechanisms of heavy quarkonia  and light hadron spectra from its decays, and
can be treated nonrelativistically and perturbatively. 
Using $(3.79\pm0.31)\times10^6 ~\psip$  sample
collected with the BEijing Spectrometer (BES) at BEPC,  
the recent $\psi'$, $\chi_c$  and $\eta_c$ 
results are presented. The BES detector is described in detail in
Ref. \cite{BES}. 
\vskip 0.1cm 
Out of the charmonium decays, there exists a mysterious and longstanding
$\rho\pi$ puzzle of $J/\psi$ and $\psip$ decays \cite{rhopi},
first revealed by Mark-II Collaboration.  In QCD, both $\psip$ and 
$\jpsi$ decays are expected to be dominated by annihilations 
into three gluons. Based on the similarity between the $\psip$ and $\jpsi$ wave
functions and the slow running of the strong interaction
coupling constant $\alpha_s$, 
the non-relativistic perturbative QCD 
predicts \cite{qcd15}: the ratios of the decays widths of $\psi'$ into hadrons 
to those of $J/\psi$ into hadrons are: 
$$Q_h=\frac{Br(\psi'\to h)}{Br(J/\psi\to h)}\simeq
\frac{Br(\psi'\rightarrow e^+e^-)}{Br(\psi\rightarrow e^+e^-)}
=(14.7\pm2.3)\%$$
named as the ``$14\%$'' rule. Mark II first observed the vector-pseudoscalar 
(V for vector, P pseudoscalar, B baryon, A axial-vector and T tensor) 
$\rho\pi$ and
$K^*\bar K$ channels are highly suppressed w.r.t. the $``14\%$'' expectation
-- known as the $\rho\pi$ puzzle. BES has confirmed the puzzle with much 
lower upper limit by a suppression factor of $\sim60$ and observed
new highly suppressed decay modes, meanwhile BES 
presents many observations (Table \ref{hadronic}).

\begin{figure}[h]
\vskip  -1.3cm
\centerline{\epsfig{figure=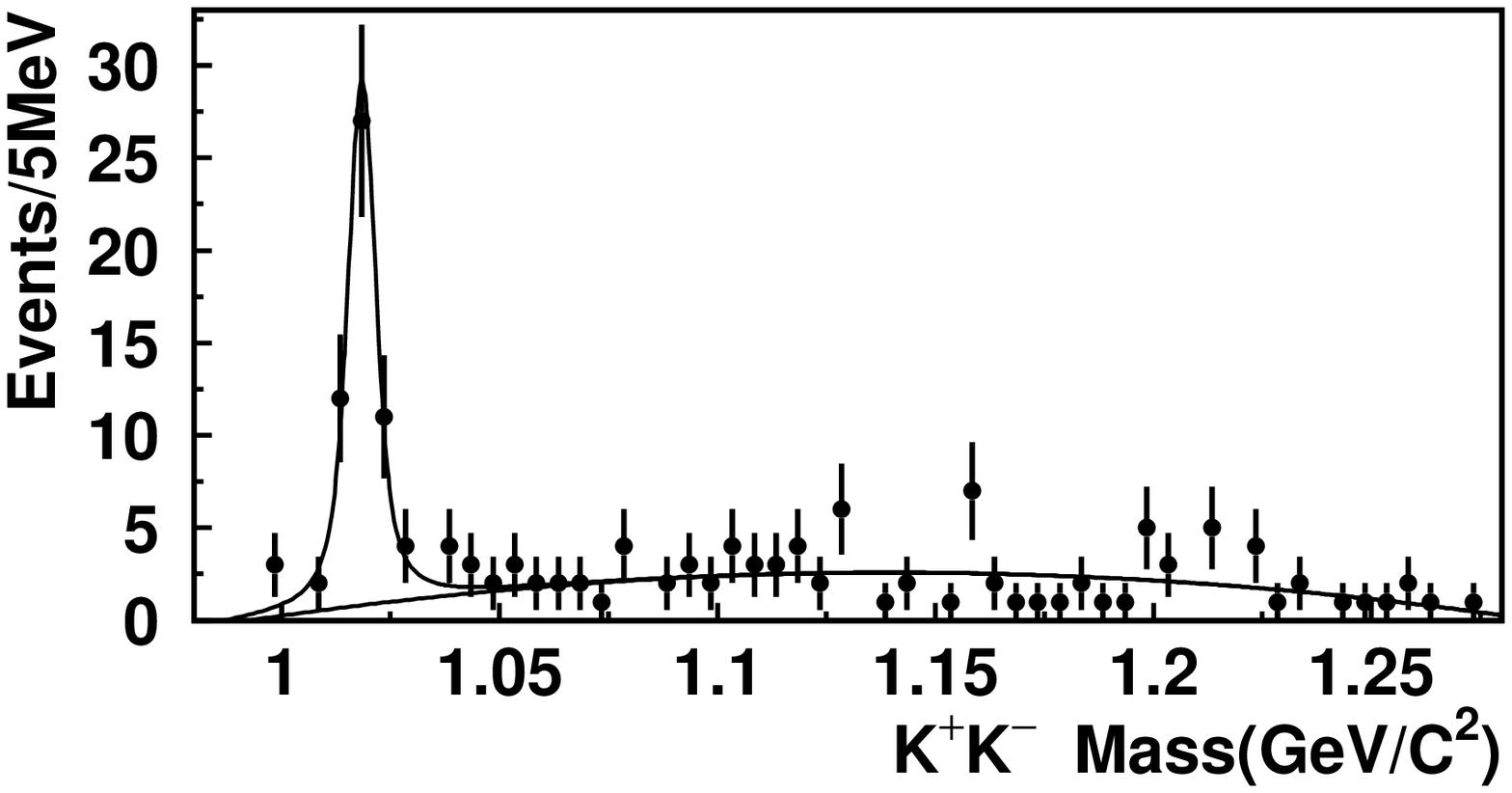,width=6.0cm}%,height=5.5cm}
\epsfig{figure=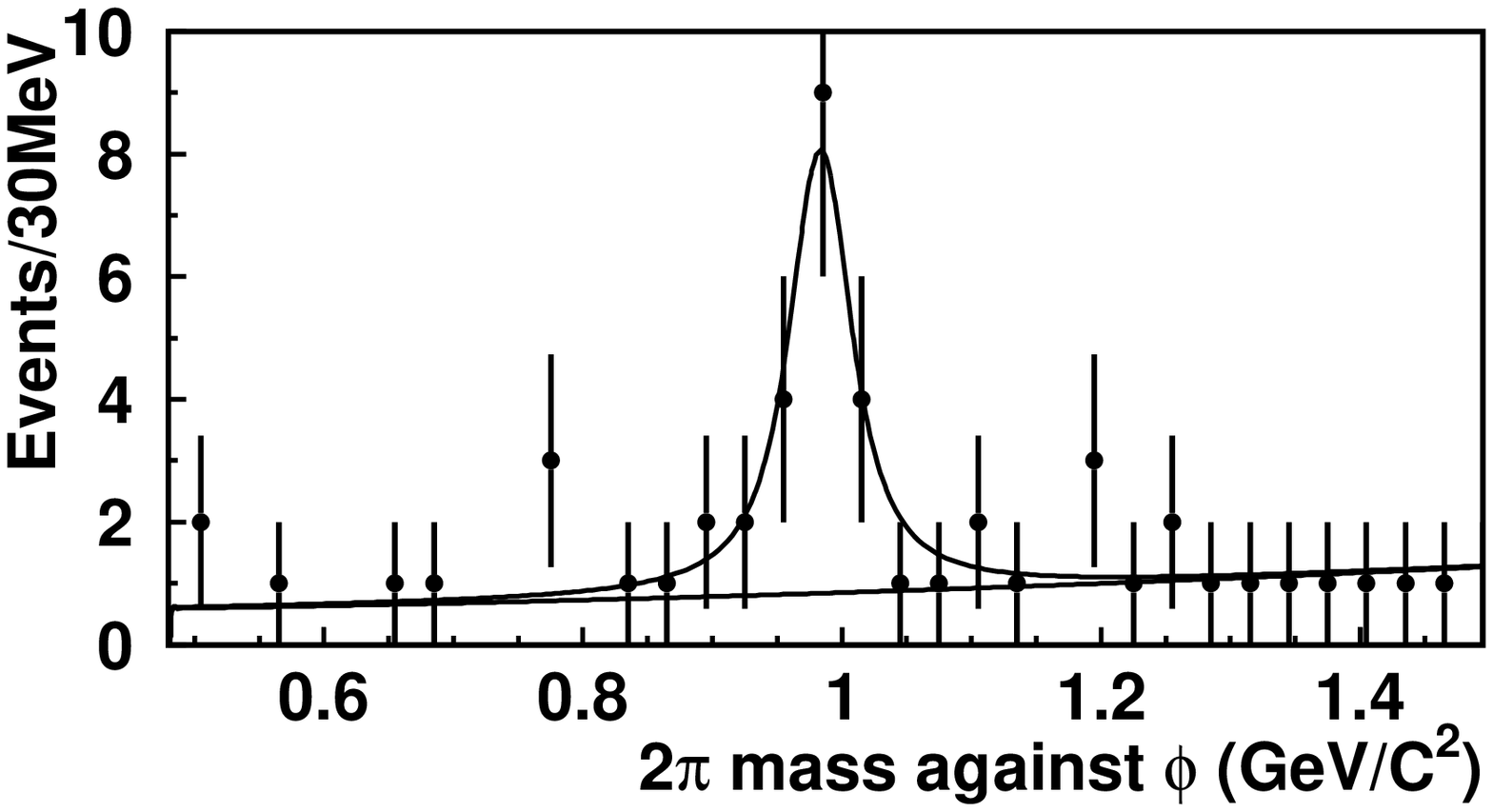,width=6.0cm}}%,height=5.5cm}}
\vskip -1.0cm
\caption[.]{$K^+K^-$ and $\pi^+\pi^-$ masses for $\psip\to\phi\pi^+\pi^-$
and $\phi f_0(980)$ (preliminary).}
\label{phif0}
\end{figure}

From Table \ref{hadronic}, a large number of $\psi'$ decay branching ratios have
been measured, and most for the first time \cite{PDG98}.
And it is found  that some VP, VPP, VT, PBB and AP
decay modes are suppressed w.r.t. the $``14\%$'' rule, where VT modes \cite{vt} 
is the first evidence for
suppression other than VP. While for $\psi'$ decays into AP modes \cite{AP},
the normal decay $b_1\pi$, enhanced mode $K_1(1270)\bar K$
and suppressed decay $K_1(1400)\bar K$ are observed. The $``14\%$'' rule
holds for the radiative decays into VT modes \cite{gpp} within errors,
but it is suppressed for the decays into VP modes\cite{radiat}.  
Now the suppressed decays also extend to the 3-body decays, 
but the $``14\%$'' rule holds for the baryonic decays \cite{baryon}. 
Quite a few models \cite{models} have been put
forward to explain the puzzle, but none of them is satisfactory.
From the measured results, it has also been observed that 
the  {\bf strong double OZI (DOZI) violation}
in $\psip\to\phi \pi^+\pi^-$  and the isospin violation in
the decays between $\psip\to\omega\pi^0$ and the isospin-conserving 
and SU(3)-allowed decay $\psip\to\rho\pi$, and between the charged and neutral 
decays $\psip\to K^{*+}K^-+c.c$ and $\psip\to K^{*0}\bar K^0+c.c$.
Fig. \ref{phif0} 
shows the $K^+K^-$ and $\pi^+\pi^-$ invariant masses 
for $\psip\to\phi\pi^+\pi^-$ and $\phi f_0(980)$ \cite{omegaphi}.  
Also BES first measures
${Br(\psi'\to\tau^+\tau^-)}=(2.82\pm0.45\pm0.56)\times10^{-3} 
~ \rm (preliminary)$
and has precisely  determined 
$Br(J/\psi\rightarrow \ell^+\ell^-)=(5.87\pm0.04\pm0.09)\%$ \cite{psill} 
 with errors half of those in PDG98. 

{\footnotesize%\tiny
\begin{table}[htb]
\noindent\vskip -1.2cm 
\caption{\label{hadronic} $\psi'$ decay branching ratios (the unpublished
results indicated by * are preliminary).}
\begin{center}
\noindent\vskip -0.0cm
\begin{tabular}{|l|c|c|c|}\hline 
Channels  &{$Br_{PDG}(\times10^{-4})$}&$Br_{BES}(\times10^{-4})$ &
$Q_h^{BES}=\frac{Br_{\psip}}{Br_{\jpsi}}(\%)$ \\ \hline
 $^* ~ \rho \pi$	& $ <0.83$  &	$ < 0.28$ & $<0.22 $\\	
 $^* ~ K^+\overline{K}^*(892)^- + c.c.$  
		& $<0.54$   &  $<0.30$ &$ <0.6 $ \\		
 $^* ~ K^0 \overline{K}^* (892)^0 + c.c.$  
		& -- &  $0.81 \pm 0.24 \pm 0.16$ & $ 1.9\pm 0.7$\\ 
 $^* ~ \omega \pi^0$	&   -- &  $0.38 \pm 0.17 \pm 0.11$ & $9.1\pm5.0$ \\	
 $^* ~ \omega \eta$&   -- &  $<0.33$ & $<2.1 $\\
 $^* ~ \pi^+\pi^-\pi^0$& 0.9$\pm$0.5 & $ 1.06\pm0.11\pm0.16$ & $7.1\pm1.6 $\\  
 $^* ~ K\overline{K}\pi$& -- & 1.25 $\pm$ 0.18 $\pm$ 0.26 &$2.1\pm0.6$\\ 
\hline\hline
 $\gamma \eta         $  &   -- & $0.53\pm0.31\pm0.08$ & $ 6.2\pm3.8$ \\
$\gamma \eta'(958)$  &   -- & 1.54 $\pm 0.31 \pm 0.20$ &$3.6\pm0.9$\\ \hline
$\omega f_2$&   -- &  $<1.7$	& $<3.9 $	    \\
$\rho a_2$	&   -- &  $<2.3$	& $<2.1$\\ 
$K^*(892)^0\overline{K}^*_2(1430)^0+c.c.$ 
    &   -- &  $< 1.2$     			&$<1.8 $ \\ 
$\phi f^\prime_2 (1525)$ &   -- & $<0.45$&$<3.66$  \\ \hline
$b_1 \pi$& -- & 5.2 $\pm$ 0.8 $\pm$ 1.0 & $17.3\pm5.2$\\
$K_1 (1270) \overline{K}$& -- &   10.0 $\pm$ 1.8 $\pm$ 2.1 &
$>33.4$ \\
$K_1 (1400) \overline{K}$& -- & $< 3.1$  & $<8.2$\\ \hline\hline
$^* ~ \gamma f_2(1270)$    &  --  &  $2.36 \pm 0.24 \pm 0.32$ & $17.1\pm2.9$\\
$^* ~ \gamma f_J(1710)$    &  --  &  $1.24 \pm 0.20 \pm 0.16$ & $12.8\pm3.4$\\ \hline
$^* ~ \phi K^+ K^-$		& -- & 0.61 $\pm$ 0.18 $\pm$ 0.15  &$7.3 \pm 3.0$ \\ 
$^* ~ \phi \pi^+ \pi^-$      & -- & 1.76 $\pm$ 0.22  $\pm 0.24$ & $ 22.0 \pm 5.3$\\ 
$^* ~ \phi f_0(980)$		& -- & 0.66 $\pm0.17\pm0.07$& $20.5\pm8.1$\\ 
$^* ~ \omega \pi^+ \pi^-$    & -- & 4.47 $\pm 0.57 \pm 0.54$   &$6.3\pm 1.4$ \\ 
$^* ~ \omega K^+ K^-$        & -- & 1.26 $\pm 0.43\pm0.39$ &$17.0\pm9.6$ \\ 
$^* ~ \omega p\bar p$ 	& -- & 0.60 $\pm0.23\pm0.14$&$4.6\pm2.2$ \\ 
$^* ~ \phi p\bar p$		& -- & 0.86 $\pm0.50\pm0.20$ &$18.9\pm13.4$ \\
$^* ~ \pi^+\pi^-\pi^0p\bar p$& -- & 3.85 $\pm0.37\pm0.56$& 16.7$\pm$7.1\\
$^* ~ \eta\pi^+\pi^-p\bar p$ & -- & 2.58 $\pm 0.68\pm0.73$ & \\ 
$^* ~ \eta p\bar p$& -- & 0.84 $\pm 0.48\pm0.48$ &4.00$\pm$4.00 \\ \hline 
$^* ~ p\bar p$ &$1.9\pm0.5$& $2.62\pm0.17\pm0.55$&$12.3\pm2.8$\\ 
$^* ~ \Lambda\overline{\Lambda}$ & $<4.0$  & 1.89 $\pm0.21\pm0.21$ & $14.0\pm2.7$\\ 
$^* ~ \Sigma^0\overline\Sigma^0$ & -- & 1.20 $\pm0.40\pm0.30$ & $9.2\pm4.1 $ \\
$^* ~ \Xi^-\overline{\Xi}^+$ & $<2.0$ & 1.00 $\pm0.30\pm0.10$ & $11.1\pm 4.3$ \\
$^* ~ \Delta^{++}\overline{\Delta}^{--}$ & -- & 1.34 $\pm0.11\pm0.33$ & $12.2\pm4.5 $\\ 
$^* ~ \Sigma^{*+}\overline\Sigma^{{*-}}$ & -- & 1.10 $\pm0.30\pm0.30$ & $10.7\pm4.3$  \\
\hline
\end{tabular}
\vskip -1.4cm 
\end{center}
\end{table}
} %end footnotesize

\section{ Hadronic $\chi_{cJ}$ Decays}

The $(3.79\pm0.31)\times10^6 ~\psip$ sample permits studies
of $\chi_{cJ}$ with
unprecedented precision ($\sim1.0\times10^6 \chi's$). 
Table \ref{chi_tbl} shows the results of $\chi_{cJ}$ 
decays \cite{chicpp,chichad}. From the table, the measurement precision is 
improved much, and many decay modes are presented first time, 
like $\chi_{c0}\to p\bar p$ \cite{chicpp}. The 
results of the neutral decays are preliminary. 
{\footnotesize%\tiny
\begin{table}[thb]
\vskip -1.0cm 
\caption{\label{chi_tbl}Results on $\chi_{cJ}$ Hadronic Decays (the 
unpublished results indicated by * are preliminary).}
\begin{center}
\vskip -0.0cm 
\begin{tabular}{|l|c|c|} \hline  
decay channels&{ BES  $ (\times 10^{-3})$} &{ PDG$(\times 10^{-3})$}\\ \hline\hline
$Br(\chi_{c0} \to \pi^+ \pi^-)$
          &    4.68 $\pm$ 0.26 $\pm$ 0.65    &   7.5 $\pm$ 2.1 \\
$Br(\chi_{c2} \to \pi^+ \pi^-)$
          &  1.49 $\pm$ 0.14 $\pm$ 0.22     &  1.9 $\pm$ 1.0   \\ \hline 
$Br(\chi_{c0} \to K^+ K^-)$
          &    5.68 $\pm$ 0.35 $\pm$ 0.85   &   7.1 $\pm$ 2.4 \\
$Br(\chi_{c2} \to K^+ K^-)$
          &  0.79 $\pm$ 0.14 $\pm$ 0.13    &  1.5 $\pm$ 1.1   \\ \hline
$Br(\chi_{c0} \to p\overline{p})$
          &    0.159 $\pm$ 0.043 $\pm$ 0.053   &   $<0.9$ \\
$Br(\chi_{c1} \to p\overline{p})$
          &  0.042 $\pm$ 0.022 $\pm$ 0.028     &  0.086 $\pm$ 0.012  \\
$Br(\chi_{c2} \to p\overline{p})$
          &  0.058 $\pm$ 0.031 $\pm$0.032& 0.10 $\pm$ 0.01\\ \hline\hline

$Br(\chi_{c0} \to \pi^+ \pi^-\pi^+ \pi^-)$
          &    15.4 $\pm$ 0.5 $\pm$ 3.7  &   37 $\pm$ 7 \\

$Br(\chi_{c1} \to \pi^+ \pi^-\pi^+ \pi^-)$
          &  4.9 $\pm$ 0.4 $\pm$ 1.2     &  16 $\pm$ 5  \\

$Br(\chi_{c2} \to \pi^+ \pi^-\pi^+ \pi^-)$
          &  9.6 $\pm$ 0.5 $\pm$ 2.4    &  22 $\pm$ 5   \\ \hline
$Br(\chi_{c0} \to K^0_s K^0_s)$
          &   1.96 $\pm$ 0.28 $\pm$ 0.52   &    -   \\
$Br(\chi_{c2} \to K^0_s K^0_s)$
          &  0.61 $\pm$ 0.17 $\pm$ 0.16 &   - \\ \hline
$Br(\chi_{c0} \to \pi^+ \pi^- K^+ K^-)$
          & 14.7 $\pm$ 0.7 $\pm$ 3.8  &   30 $\pm 7$ \\

$Br(\chi_{c1} \to \pi^+ \pi^- K^+ K^-)$
          & 4.5 $\pm$ 0.4 $\pm$ 1.1 &   9 $\pm 4$ \\

$Br(\chi_{c2} \to \pi^+ \pi^- K^+ K^-)$
          & 7.9 $\pm$ 0.6 $\pm$ 2.1  &   19 $\pm 5$ \\ \hline

$Br(\chi_{c0} \to \pi^+ \pi^- p \bar{p})$
          & 1.57 $\pm$ 0.21 $\pm$ 0.54   &   5.0 $\pm 2.0$ \\
$Br(\chi_{c1} \to \pi^+ \pi^- p \bar{p})$
          & 0.49 $\pm$ 0.13 $\pm$ 0.17  &   1.4 $\pm 0.9$ \\
$Br(\chi_{c2} \to \pi^+ \pi^- p \bar{p})$
          & 1.23 $\pm$ 0.20 $\pm$ 0.35  &   -- \\  \hline

$Br(\chi_{c0} \to K^+ K^- K^+ K^-)$
        &  2.14 $ \pm$ 0.26 $\pm$ 0.40  &    --  \\

$Br(\chi_{c1} \to K^+ K^- K^+ K^-)$
        &  0.42 $ \pm$ 0.15 $\pm$ 0.12 &    --  \\

$Br(\chi_{c2} \to K^+ K^- K^+ K^-)$
        &  1.48 $ \pm$ 0.26 $\pm$ 0.32 &    --  \\  \hline

$Br(\chi_{c0} \to \phi  \phi)$
        &   0.92   $\pm$ 0.34 $\pm$ 0.38  &  --  \\

$Br(\chi_{c2} \to \phi  \phi)$
        &   2.00   $\pm$ 0.55 $\pm$ 0.61  &  --  \\  \hline

$Br(\chi_{c0} \to K^0_s K^+ \pi^- + c.c.)$
        &   $< 0.71$     &  --  \\

$Br(\chi_{c1} \to K^0_s K^+ \pi^- + c.c.)$
        &   $2.46 \pm 0.44 \pm 0.65$    &  --  \\

$Br(\chi_{c2} \to K^0_s K^+ \pi^- + c.c.)$
        &   $<1.06 $     &  --  \\ \hline
$Br(\chi_{c0} \to 3(\pi^+ \pi^-))$
        &   $11.7 \pm 1.0 \pm 2.3$    &  $15 \pm 5$  \\
$Br(\chi_{c1} \to 3(\pi^+ \pi^-))$
        &   $5.8 \pm 0.7 \pm 1.2$    &  $22 \pm 8$  \\
$Br(\chi_{c2} \to 3(\pi^+ \pi^-))$
        &   $9.0 \pm 1.0  \pm 2.0$    &  $12 \pm 8$  \\ \hline\hline

$^*Br(\chi_{c0} \to \pi^0 \pi^0)$
          &    2.80 $\pm$ 0.32 $\pm$ 0.51    &  $3.1\pm0.6$  \\
$^*Br(\chi_{c2} \to \pi^0 \pi^0)$
          &  0.92 $\pm$ 0.27 $\pm$ 0.52     &  $1.1\pm0.3$   \\ \hline
$^*Br(\chi_{c0} \to \eta \eta)$
          &  2.03 $\pm$ 0.84 $\pm$ 0.58     & $2.5\pm1.1$   \\ 
$^*Br(\chi_{c2} \to \eta \eta)$ &   $<2.5$    &  $0.8\pm0.5$  \\ \hline
\end{tabular}
\vskip -1.2cm 
\end{center}
\end{table}}

\begin{figure}[h]
\vskip  -2.0cm
\centerline{\epsfig{figure=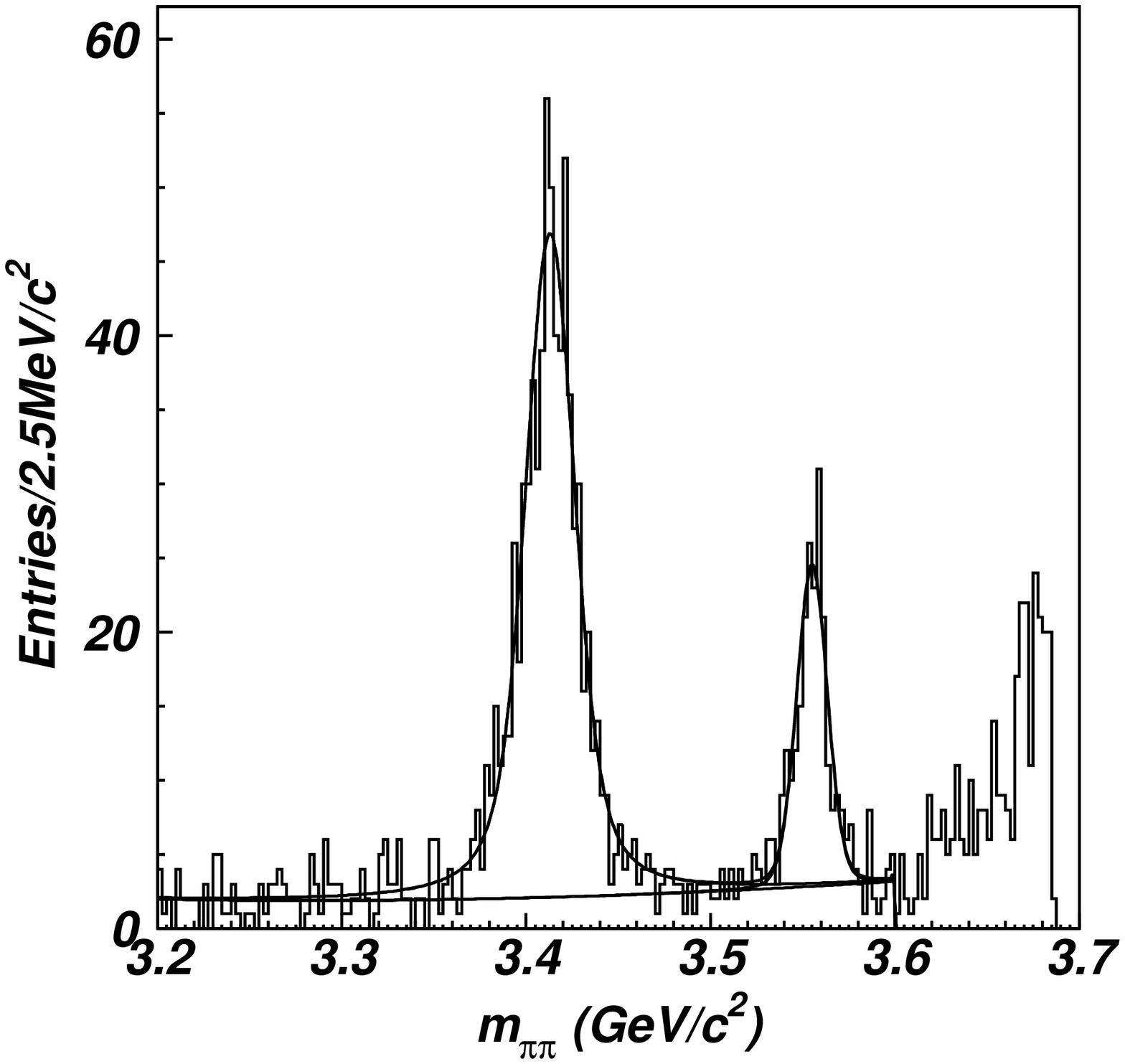,width=3.8cm}
\epsfig{figure=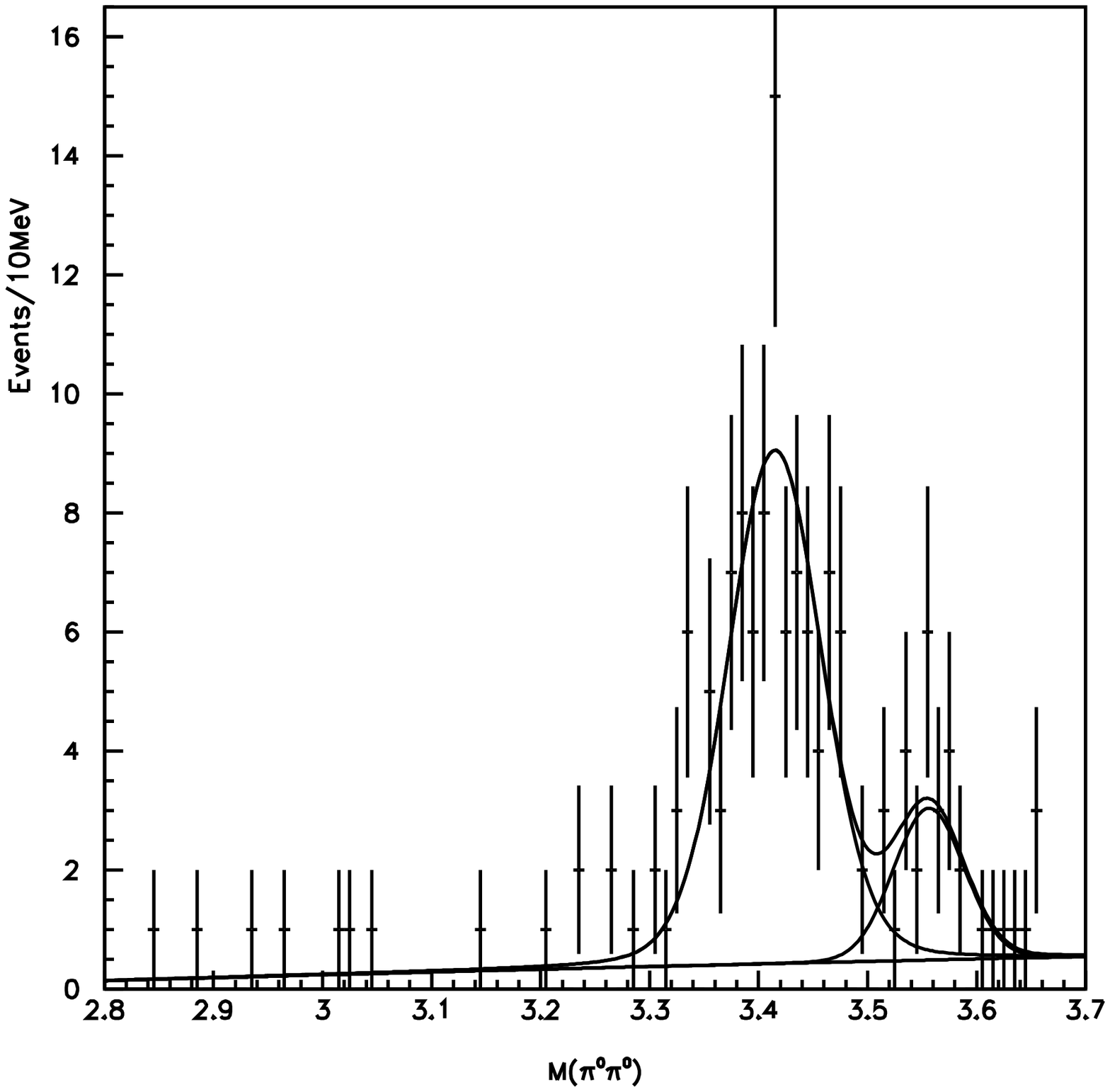,width=3.8cm} 
\epsfig{figure=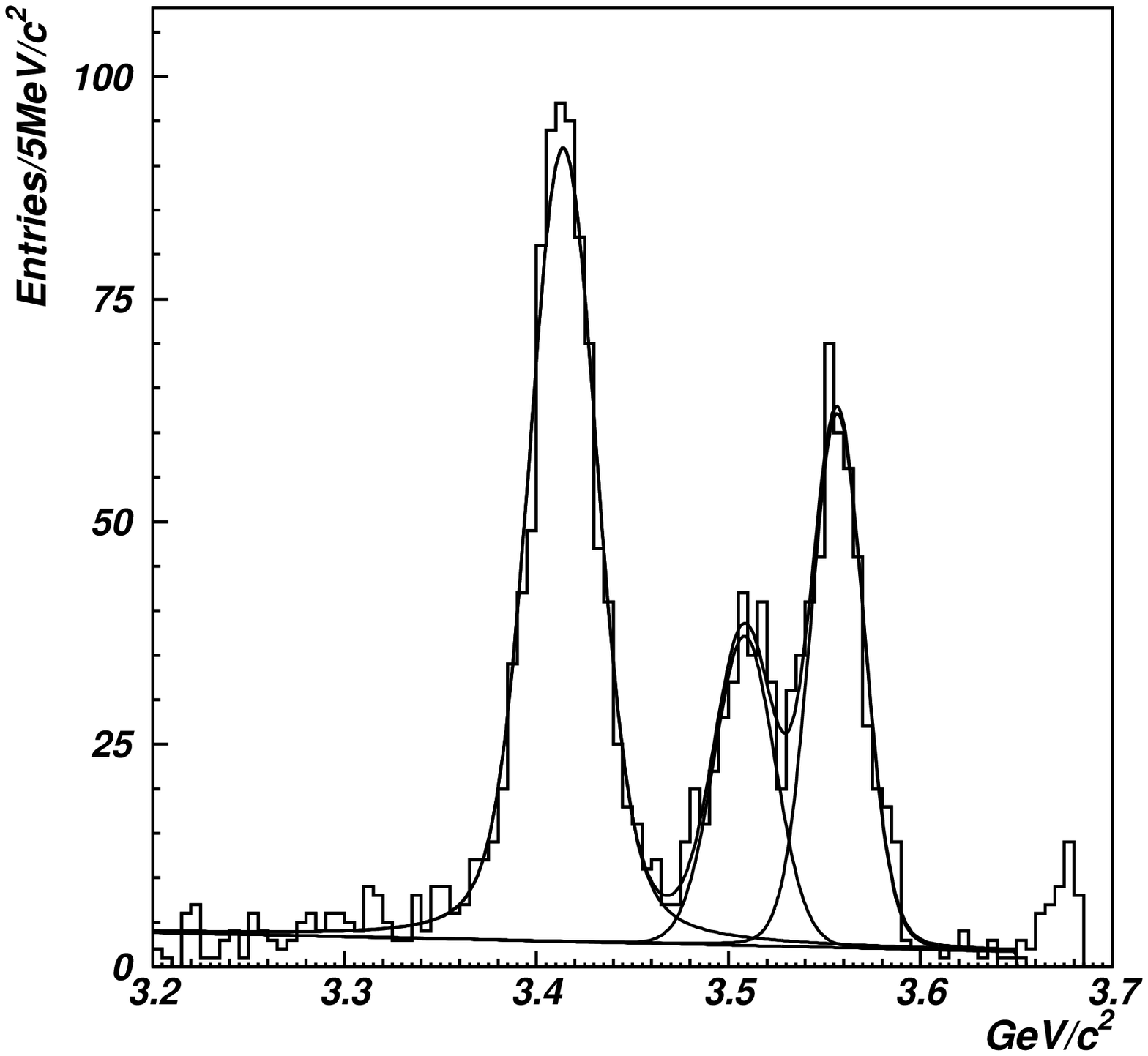,width=3.8cm}
\epsfig{figure=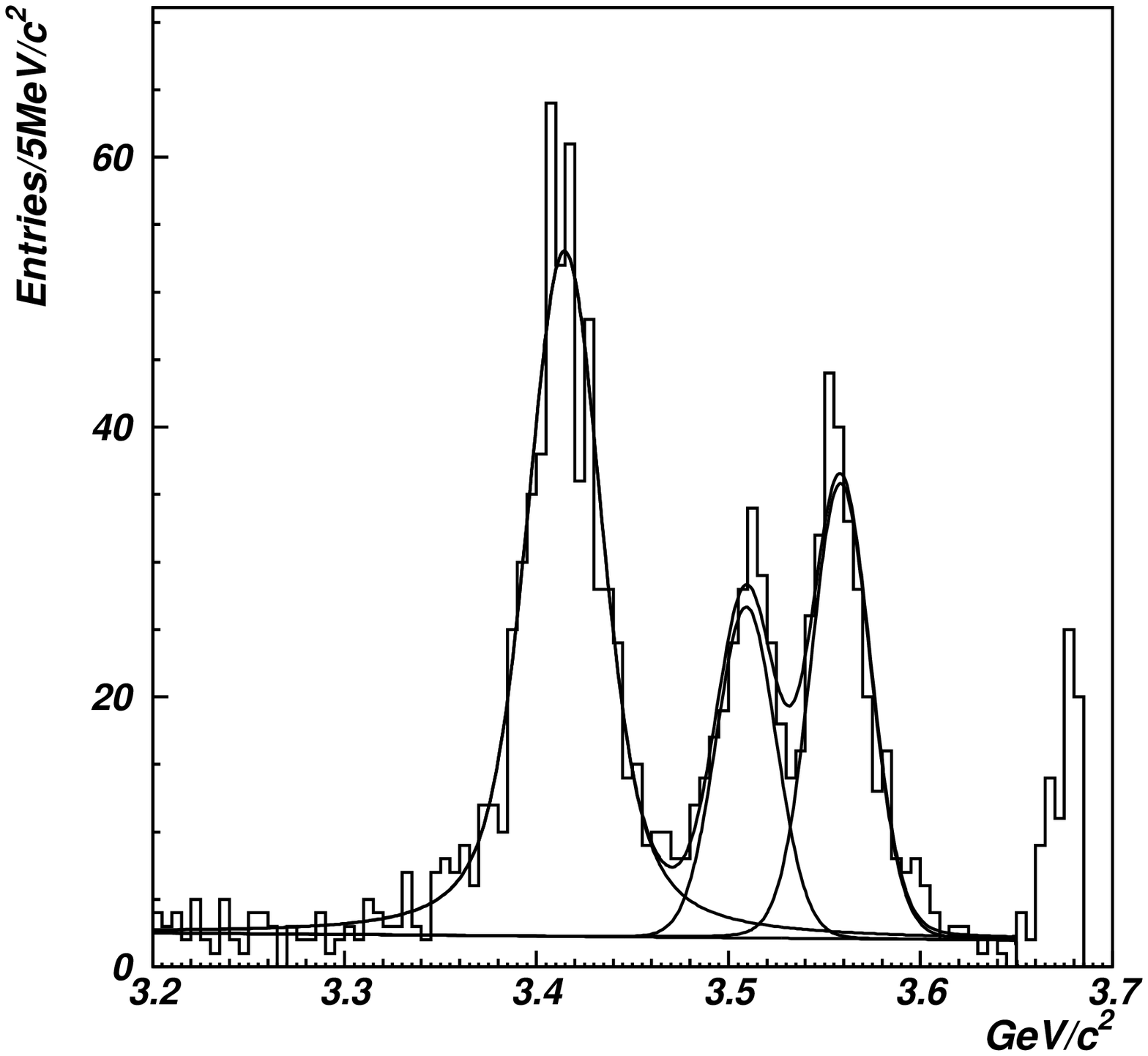,width=3.8cm}}
\vskip -1.1cm
\caption[.]{$\chi_{cJ}$ decays into $\pi^+\pi^-$ \cite{chicpp},
 $\pi^0\pi^0$ (preliminary), $2(\pi^+\pi^-)$ and 
 $\pi^+\pi^- K^+K^-$ \cite{chichad} respectively.}
\label{chi2pi}
\end{figure}

The mass differences between $\chi_{c0,1,2}$ and between $\eta_c$ and $J/\psi$
reflect the hyperfine structure of the spin-spin interactions, 
the $\chi_{c1,2}$ and $J/\psi$ masses have been precisely 
determined.  BES improves \cite{chicpp} the $\chi_{c0}$ 
mass with ($3414.1\pm0.6\pm0.8$) MeV over PDG98 value
with an error of 2.8 MeV, and $\chi_{c0}$ decay width with
$\Gamma_{\chi_{c0}}=(14.3\pm3.6)$ MeV over PDG98 value ($13.5\pm5.3$ MeV).  
Fig. \ref{chi2pi} 
shows $\chi_{cJ}$ hadronic decays 
\cite{chicpp,chichad}. 
From $\psi'$ radiative transition to $\eta_c$, BES gives the $\eta_c$ mass 
with ($2975.8\pm3.9\pm1.2$) MeV \cite{chichad}.
\vskip 0.2cm 
\section{Comparison with NRQCD}

Recent years, non-relativistic QCD (NRQCD) \cite{octet} are successfully 
applied into description of the production and decays of heavy quarkonia.
Its key idea is that heavy quark $Q\bar Q$
pairs are produced at short distances
in color-octet states and subsequently evolve into physical (color-singlet)
quarkonia by nonperturbative emission of soft gluons. Using the total decay 
widths of $J/\psi$, $\psi'$ and $\chi_{c0,1,2}$ from
PDG98 and the BES measured branching fractions, the 
corresponding partial decay widths are extracted,  see Table \ref{NRQCD}. From 
Table \ref{NRQCD}, it is shown that the experimental results accord with NRQCD 
predictions \cite{octet2} within errors. But for some decay modes, 
the errors are large due to the low statistic. 
{\footnotesize%\tiny
\begin{table}[h]
\noindent\vskip -0.7cm 
\caption{\label{NRQCD} Charmonium Partial Decay Decay Widths.} 
\begin{center}
\begin{tabular}{|l|c|c|c|}\hline 
Channels  & { NRQCD} & PDG &{ BES }\\ \hline\hline
$J/\psi\to p\bar p$              &{ 174 eV}  & ($186\pm14$) eV&--  \\
$J/\psi\to \Sigma^0\bar\Sigma^0$ & { 113 eV} & ($110\pm16$) eV&--   \\
$J/\psi\to \Lambda\bar\Lambda $  & { 117 eV} & ($117\pm14$) eV&--  \\
$J/\psi\to \Xi^-\bar\Xi^+$       &{  62.5 eV} & ($78\pm18$) eV&--  \\
$J/\psi\to \Delta^{++}\overline{\Delta}^{--}$ & { 65.1 eV} & ($96\pm26$) eV&--  \\
$J/\psi\to \Sigma^{*-}\bar\Sigma^{*+}$ & { 40.8 eV} & ($45\pm6$) eV &--   \\ \hline\hline
$\psi'\to p\bar p$ &{ 76.8 eV} &($52.6\pm15.1$) eV&{ ($72.6\pm18.0$) eV}  \\
$\psi'\to \Sigma^0\bar\Sigma^0$&{ 55.0 eV} &--&{  ($33.2\pm14.3$) eV}   \\
$\psi'\to \Lambda\bar\Lambda $&{ 54.6 eV} &--&{ ($52.4\pm10.2$) eV}  \\
$\psi'\to \Xi^-\bar\Xi^+$   & { 33.9 eV} &--&{ ($27.7\pm9.4$) eV}  \\
$\psi'\to \Delta^{++}\overline{\Delta}^{--}$ & { 32.1 eV} &--&{ ($37.2\pm10.5$) eV}  \\
$\psi'\to \Sigma^{*-}\bar\Sigma^{*+}$ &{  24.4 eV} &--&{ ($30.5\pm12.1$) eV}   \\ \hline\hline
$\chi_{c1}\to p\bar p$  &{ 56.2 eV}  &($75.7\pm16.0$) eV &{ ($37.0\pm32.2$) eV}  \\
$\chi_{c2}\to p\bar p$  &{ 154.2 eV} &($200.0\pm27.0$) eV &{  ($116.0\pm90.6$) eV} \\
$\chi_{c0}\to K^+K^-$       &{ 38.6 keV }&($99\pm49$) keV &{ ($79.5\pm31.2$) keV}\\ 
$\chi_{c2}\to K^+K^-$       &{ 2.89 keV }&($3.0\pm2.2$) keV &{ ($1.58\pm0.41$) keV}\\ 
$\chi_{c0}\to \pi^+\pi^-$   &{ 45.4 keV}&($105\pm48$) keV &{($65.5\pm25.4$) keV}\\ 
$\chi_{c2}\to \pi^+\pi^-$   &{ 3.64 keV }&($3.8\pm2.0$) keV&{ ($2.98\pm0.59$) keV}\\ \hline\hline
$\chi_{c0}\to \pi^0\pi^0$   &{ 23.5 keV }&($43\pm18$) keV  &{  ($39.2\pm16.3$) keV}\\ 
$\chi_{c2}\to \pi^0\pi^0$   &{ 1.93 keV }&($2.2\pm0.6$) keV &{ ($1.84\pm1.20$) keV}\\ 
$\chi_{c0}\to\eta\eta$      &{ 24.0 keV} &($35.0\pm20.2$) keV&{ ($28.4\pm17.5$) keV }\\
$\chi_{c2}\to\eta\eta$      &{ 1.91 keV }&($1.6\pm1.0$) keV &{ $<5$ keV} \\     
\hline
\end{tabular}
\vskip -1.2cm 
\end{center}
\end{table}}

\section{Summary}

Using the $(3.79\pm0.31)\times10^6$ $\psi'$ data sets at BES, 
a large number of $\psip$ and $\chi_{cJ}$ results have been 
presented, many of them 
for the { first} measurements and/or with the
{ unprecedented precision.} BES has observed {new
suppressed} decay modes and { first
observed enhanced} decay from $\psi'$ among a large number of
the normal hadronic decays. Some results are also 
compared with NRQCD calculations. 

{\bf   Acknowledgements} 

 We acknowledge the strong efforts of the BEPC staff and the
 helpful service from the IHEP
 computing center. The work of the BES Collaboration is supported in part by
 the National Natural Science Foundation of China
 under Contract No. 19991480 and the Chinese Academy of Sciences
 under contract KJ95T-03, %No. H-10 and E-01 (IHEP),
 and by the Department of
 Energy under Contract Nos. DE-FG03-92ER40701 (Caltech),
 DE-FG03-93ER40788 (Colorado State University), DE-AC03-76SF00515 (SLAC),
 DE-FG03-91ER40679 (UC Irvine), DE-FG03-94ER40833 (U Hawaii),
 DE-FG03-95ER40925 (UT Dallas).


\begin{thebibliography}{99}
\bibitem{BES}J. Z. Bai et al., BES Coll., {\Journal\NIMA&344&319(1994).}
\bibitem{rhopi}M. E. B. Franklin et al., {\Journal\PRL&51&963(1983).}
\bibitem{qcd15}T. Applequist and D. Politzer {\Journal\PRL&34&43(1975).}
\bibitem{PDG98}Particle Data Group, R. M. Barnet et al., Euro. Phys. J. {\bf C3}
1998. 
\bibitem{vt}J. Z. Bai et al., BES Coll., {\Journal\PRL&81&5080(1998)}.
\bibitem{AP}{J. Z. Bai \etal, BES Coll.,} {\Journal\PRL&83&1918(1999)}.
\bibitem{gpp}{J. Z. Bai \etal, BES Coll.,} $\psip$ ratiative decays 
into two pseudoscalars, to be submitted. 
\bibitem{radiat}{J. Z. Bai \etal, BES Coll.,}{\Journal\PRD&58&097101(1998).}
\bibitem{baryon}{J. Z. Bai \etal, BES Coll.,} $\psip$ decays into 
baryon-antibaryon pairs, to be submitted. 
\bibitem{models}for example, W. S. Hou and A. Soni, 
Phys. Rev. Lett {\bf 50}, 569(1983);
	%{\Journal\PRL&50&569(1983)};
%	G. Karl and W. Roberts, {\Journal\PLB&144&243(1984)};
	S. J. Brodsky, G. P. Lepage and S. F. Tuan,
	Phys. Rev. Lett {\bf 59}, 631(1987);
	%{\Journal\PRL&59&631(1987)};
%	M. Chaichian et al., {\Journal\NPB&323&75(1989)};
%	S. S. Pinsky, {\Journal\PLB&236&479(1990)};
	X. Q. Li et al., Phys. Rev. {\bf D55}, 1241(1997)
	%{\Journal\PRD&55&1241(1997)};
	S. J. Brodsky and M. Karliner, Phys. Rev. Lett {\bf 78},468(1997);
	%{\Journal\PRL&78&468(1997)};
	Y. Q. Chen and E. Braaten, Phys. Rev. Lett {\bf 80},5060(1998).
	%{\Journal\PRL&80&5060(1998)}.
\bibitem{omegaphi}{J. Z. Bai \etal, BES Coll.,} $\psip$ hadronic 
decays involving $\omega/\phi$,  to be submitted. 
\bibitem{psill}{J. Z. Bai \etal, BES Coll.,}{\Journal\PRD&58&092006(1998)}.
\bibitem{chicpp}{J. Z. Bai \etal, BES Coll.,} {\Journal\PRL&81&3091(1998)}
\bibitem{chichad}{J. Z. Bai \etal, BES Coll.,} {\Journal\PRD&60&072001(1999).}
\bibitem{octet}G. T. Bodwin, E. Braaten and G. P. Lepage 
Phys. Rev.{\bf D46}, 1914(1992); ibid {\bf D51}, 1125(1995). 
\bibitem{octet2}J. Bolz, P. Kroll and G.A. Schuler, 
{\Journal\PLB&392&198(1997).}, 
Eur. Phys. J. {\bf C2}, 705 (1998); P. Kroll, hep-ph/9709393; 
S. M. H. Wong, hep-ph/9903236. 
\end{thebibliography}
\end{document}